\documentclass{aa}
\usepackage{txfonts}
\usepackage{graphicx}

\begin{document}

\title{Multiple populations of H$\beta$ emission line stars in the Large
Magellanic Cloud cluster NGC\,1971}

\author{Andr\'es E. Piatti\inst{1,2}\thanks{\email{andres.piatti@unc.edu.ar}}}

\institute{Instituto Interdisciplinario de Ciencias B\'asicas (ICB), CONICET-UNCUYO, Padre J. Contreras 1300, M5502JMA, Mendoza, Argentina;
\and Consejo Nacional de Investigaciones Cient\'{\i}ficas y T\'ecnicas (CONICET), Godoy Cruz 2290, C1425FQB,  Buenos Aires, Argentina\\
}

\date{Received / Accepted}

\abstract{We revisited the young Large Magellanic Cloud star cluster NGC\,1971 with the aim of providing
additional clues to our understanding of its observed extended Main Sequence turnoff (eMSTO),  a feature common seen in young stars clusters,which was
recently argued to be caused by a real age spread similar to the cluster age ($\sim$160 Myr). We 
combined accurate Washington and Str\"omgren photometry of high membership probability stars to explore 
the nature of such an eMSTO.  From different {\it ad hoc} defined pseudo colors we found that bluer and 
redder stars distributed throughout the eMSTO do not show any inhomogeneities of light and heavy-element
abundances. These 'blue' and 'red' stars split into two clearly different groups only when the Washington
$M$ magnitudes are employed, which delimites the number of spectral features responsible for the appearance
of the eMSTO. We speculate that Be stars populate the eMSTO of NGC\,1971 because: i) H$\beta$ 
contributes to the $M$ passband; ii) H$\beta$ emissions are common features of Be stars and; iii)
Washington $M$ and $T_1$ magnitudes show a tight correlation; the latter measuring the observed 
contribution of H$\alpha$ emission line in Be stars, which in turn correlates with H$\beta$ emissions.
As far as we are aware, this is the first observational result pointing to H$\beta$ emissions as the origin
of eMSTOs observed in young star clusters. The presence outcome will certainly open new
possibilities of studying eMSTO from photometric systems with passbands centered at features
commonly seen in Be stars.}
 

\titlerunning{Multiple populations of H$\beta$ rotators in NGC\,1971}
\authorrunning{Andr\' es E. Piatti }

\maketitle

\markboth{Andr\' es E. Piatti: Multiple populations of H$\beta$ rotators in NGC\,1971}{}

\section{Introduction} 

\citet{pc2017} detected an extended Main Sequence Turnoff (eMSTO) in the
young Large Magellanic Cloud  (LMC) cluster NGC\,1971 ($\sim$160 Myr) using broadband 
Washington $CT_1$  photometry \citep{c76}. After considering photometric uncertainties, presence
of binary stars, variations in the overall metallicity and stellar rotation effects, they
concluded that the observed eMSTO could have been caused by a real age spread of $\sim$ 170 Myr,
in addition to the above sources of color dispersion in the cluster color-magnitude diagram
(CMD). Because H$\alpha$ contributes to the Washington $T_1$ passband, the broadness of the
$C-T_1$ color range in the NGC\,1971's CMD could be due to the presence of 
Be stars, so that the eMSTO would have its origin in two populations of slow and fast
rotators, respectively. However, the $M$ vs $C-M$ CMD in \citet[][see their figure 4]{pc2017} 
still shows  an eMSTO, thus giving support to the idea of a real age spread. 
eMSTOs have been commonly seen in star clusters with ages similar to that of NGC\,1971, including Magellanic Cloud star clusters \citep[e.g.,][]{dantonaetal2017,miloneetal2017,miloneetal2018,goudfrooijetal2018}
and Galactic open clusters \citep{marinoetal2018,cordonietal2018}.

In recent years, stellar rotation has been suggested as a main source for the occurrence
of the eMSTO phenomenon in young star clusters. \citet{bastianetal2017} found 
a high fraction of Be stars in the LMC clusters NGC\,1850 ($\sim$80 Myr) and NGC\,1856 
($\sim$280 Myr) that implies a high fraction of rapidly rotating stars. These Be stars are located
toward the redder end of the cluster eMSTOs \citep[see, also][]{miloneetal2018}. \citet{dupreeetal2017}
provided with the first spectroscopic evidence that two populations coexist in the LMC
cluster NGC\,1866  ($\sim$200 Myr), consisting in one younger and slowly rotating and 
another rapidly rotating groups of stars \citep[see, also][]{gossageetal2019}. The latter exhibit 
H$\alpha$ emission lines.  Additional spectroscopic evidence is shown by 
\citet[][see, figure 2]{marinoetal2018} and \citet[][see, figure 8]{marinoetal2018b}. We note also that the effects of braking in main sequence
stars could mimic an age spread \citep{dantonaetal2017}.

In this work, we revisited NGC\,1971 with the aim of providing additional evidence on the
possible origin of its eMSTO. In Section 2 we present Str\"omgren photometry \citep{cm1976} for the
cluster field, while in Section 3 we analyze light-element abundance variations -- usually
associated to the existence of multiple populations in globular clusters \citep{marinoetal2019} --
and show that the presence of Be stars with H$\beta$ emission could have contributed to the
$M$ Washington passband and hence to the spread in the $C-M$ color. We note that
Be stars in eMSTO clusters have been detected from their emission in H$\alpha$,
no evidence for other emission features have been reported. Finally, in Section
4 we summarize the main conclusions of this work.

\section{Str\"omgren photometric data}

We searched the National  Optical Astronomy Observatory (NOAO)  Science Data Management 
(SDM)  Archives\footnote{http //www.noao.edu/sdm/archives.php.} looking for unexploited
images in the field of NGC\,1971. The observing program SO2008B-0917 (PI: Pietrzy\'nski), 
carried out with the SOAR Optical Imager (SOI) attached to the 4.1m Southern
Astrophysical Research (SOAR) telescope (FOV = 5.25$\arcmin$$\times$5.25$\arcmin$, scale=
0.154$\arcsec$/px in binned mode), obtained Str\"omgren $vby$ images centered on the
cluster during the night of January 18, 2009 under excellent image quality conditions
(typical FWHM $\sim$ 0.6$\arcsec$). They consist in 3$\times$400 sec, 3$\times$180 sec,
and 3$\times$100 sec exposures in the $v,b,y$ passbands, respectively, at airmass between 1.42 and
1.50. Standard stars were also observed during the night, namely: HD64, HD3417, HD12756, 
HD22610, HD57568, HD58489, TYC 7583-1622-1, and TYC 8104-969-1 \citep{hm1998,p2005}.
They were observed twice at a fixed airmass to allow them to be placed in the two different
CCDs arrayed by SOI, and thus to monitor their  individual responses. We also downloaded 
calibration frames (zeros, sky- and dome-flats).
We processed the images following the SOI's reduction recipes available at 
http://www.ctio.noao.edu/soar/content/soar-optical-imager-soi. The LMC cluster NGC\,1978 
was also observed during the same night, so that we took advantage of the transformation equations
fitted by \citet{pb2019}, who showed that magnitudes of standard stars placed in each CCD are 
indistinguishable. 

The image processing packages {\sc daophot}, {\sc allstar}, {\sc daomatch} and {\sc daomaster} 
\citep[stand-alone version,][]{setal90} were employed to obtain point-spread-function (PSF) 
magnitudes and their associated uncertainties. We started by selecting interactively nearly one 
hundred relatively bright, not-saturated and well-isolated stars located throughout the whole image
area to construct the corresponding PSF. A nearly 40$\%$ of the selected PSF stars were initially 
chosen to build a preliminary PSF, which was applied to the image in order to clean the whole
sample of PSF stars from fainter neighbors. We built the final quadratically 
spatially-varying PSF for that image from that cleaned PSF star sample, and computed aperture 
corrections that resulted in the range -0.04 - -0.07 mag. We obtained PSF magnitudes for the
entire list of sources identifed in the image by applying the respective PSF. From the
resulting subtracted image, we identified new sources, which were added to the previous
list to obtain simultaneously new magnitudes. We enlarged the list of sources by iterating
this procedure three times. With the aim of dealing with stellar sources we only kept those with
$\chi$ $<$ 2 and {\sc $|$sharp$|$} $<$ 0.5.  As for the mean magnitudes and their uncertainties
we straightforwardly averaged the magnitudes measured from the three images collected per
passband. Errors were estimated from extensive artificial star tests as previously performed for other subsets of Magellanic Cloud clusters imaged during the same observing program \citep{pk2018,p18b,pb2019,piattietal2019c,piatti2020}.

We followed the procedure applied by \citet{pc2017} to clean the cluster CMDs $V$ vs. $b-y$
and $V$ vs. $m_1$\footnote{$m_1 = (v-b) - (b-y)$} from field star contamination
 \citep{pb12}. The method 
relies on the use of each pair of magnitude and color in a reference field star CMD to subtract
the closest star in the cluster CMD. In  doing this, we considered the uncertainties in magnitudes 
and colors by repeating the procedure hundreds of times with magnitudes and colors varying within
their respective errors. We used as reference fields four different regions with areas equal to the
cluster region, which was delimited by a radius of 20 arcsec from the cluster center. In the end, 
we subtracted a number of stars from the cluster CMD equal to that in each reference field star 
CMD. From the four resulting cleaned cluster CMDs we assigned  membership probabilities 
to each stars based on the number of times they were subtracted in the four cleaning
procedures (one per reference field CMD). For instance, stars that appear in all the cleaned
CMDs have a membership probability $P$ = 100\%, while those seen in three cleaned
CMDs, $P$ = 75\%. We tried to complement our membership probability assignments with kinematical
information from the {\it Gaia} DR2 database \citep{gaiaetal2016,gaiaetal2018b}. Unfortunately,
the relatively small and crowded cluster field and its relatively low brightness made 
in practice unfeasible any accurate measurement of proper motion and parallax of cluster
stars. We barely got  a ten of stars located within the cluster radius with very distinct 
proper motion values.

The Washington $CMT_1$ photometry published by \citet{pc2017} was cross-matched with the
present Str\"omgren photometry, and a master table that included the membership
probabilities assigned independently from these data sets was built. In the subsequent analysis, we
did not use the $T_1$ magnitudes, which could be affected by contributions from H$\alpha$
emission lines, so that the hypothesis of the presence of slow and fast rotators as sources of the 
eMSTO was avoided from our analysis. Nevertheless, the combination of the Washington $CM$ and
Str\"omgren $vby$ magnitudes allowed us to play with different pseudo colors 
[$\equiv$($m_1-m_2$) - ($m_2-m_3$), with $m_1,m_2,m_3$ being three different magnitudes]
to uncover possible trails of the observed eMSTO in NGC\,1971.  We refer the reader to
\citet{monellietal2013} and \citet{miloneetal2017} for the definition and usefulness of
pseudo colors in the context of multiple population analyses.

\section{Analysis and discussion}

\begin{figure}
\includegraphics[width=\columnwidth]{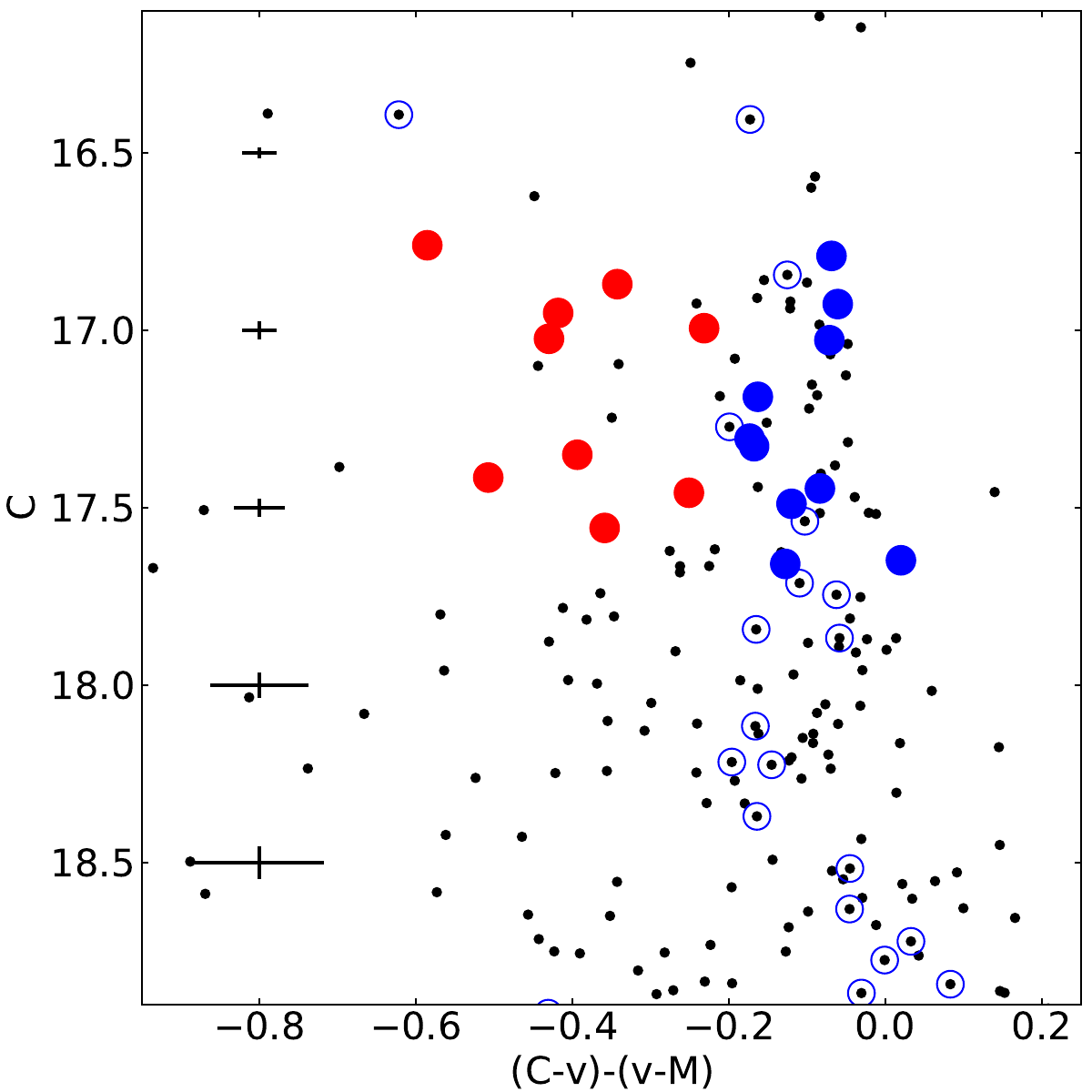}
\caption{CMD for  all the stars measured in the field of NGC\,1971 (black dots). Blue and red filled circles represent
stars with membership probabilities $P$$\ge$ 75\% in the Washington and Str\"ongren data sets,
respectively,  and located nearly covering the cluster eMSTO (16.7 $<$ $C$ (mag) $<$ 17.7). Open symbols represent all the stars with $P$$\ge$50\%. 
}
\label{fig:fig1}
\end{figure}

\citet{pc2017} showed that theoretical isochrones for the cluster age and overall metallicities between
[Fe/H] = -0.6 and +0.1 dex (the cluster metallicity is [Fe/H] = -0.3 dex \citep{dg2000}) do not account
for the observed eMSTO. However, light-element abundance variations were not sought. Indeed, 
the Washington $C$ and Str\"omgren $v$  magnitudes could reflect CN/CH variations
\citep{cummingsetal2017,limetal2017}. The $C$ and $v$ passbands have effective wavelengths
(and FWHMs) at $\lambda$3910\AA\, (1100\AA) and $\lambda$416\AA\,(190\AA), respectively, so that 
both include the CN absorption band at $\lambda$4142\AA; the $C$ passband also includes the
violet CN bands  at $\lambda$3595\AA\, and $\lambda$3883\AA. We note that the $M$ bandpass
($\lambda$5058\AA\, (1050\AA)) is freed from CN absorption bands. Figure~\ref{fig:fig1} shows
the $C$ vs. ($C-v$)-($v-M$) diagram where we highlighted with blue and red filled circles stars
that belong to the eMSTO, distributed from the top of the cluster MS down to one magnitude underneath, and with  ($C-v$)-($v-M$) values higher and lower than -0.20 mag, respectively.
These stars have membership probabilities $\ge$ 75\% in the Washington and Str\"omgren data 
sets, respectively. 

\begin{figure}
\includegraphics[width=\columnwidth]{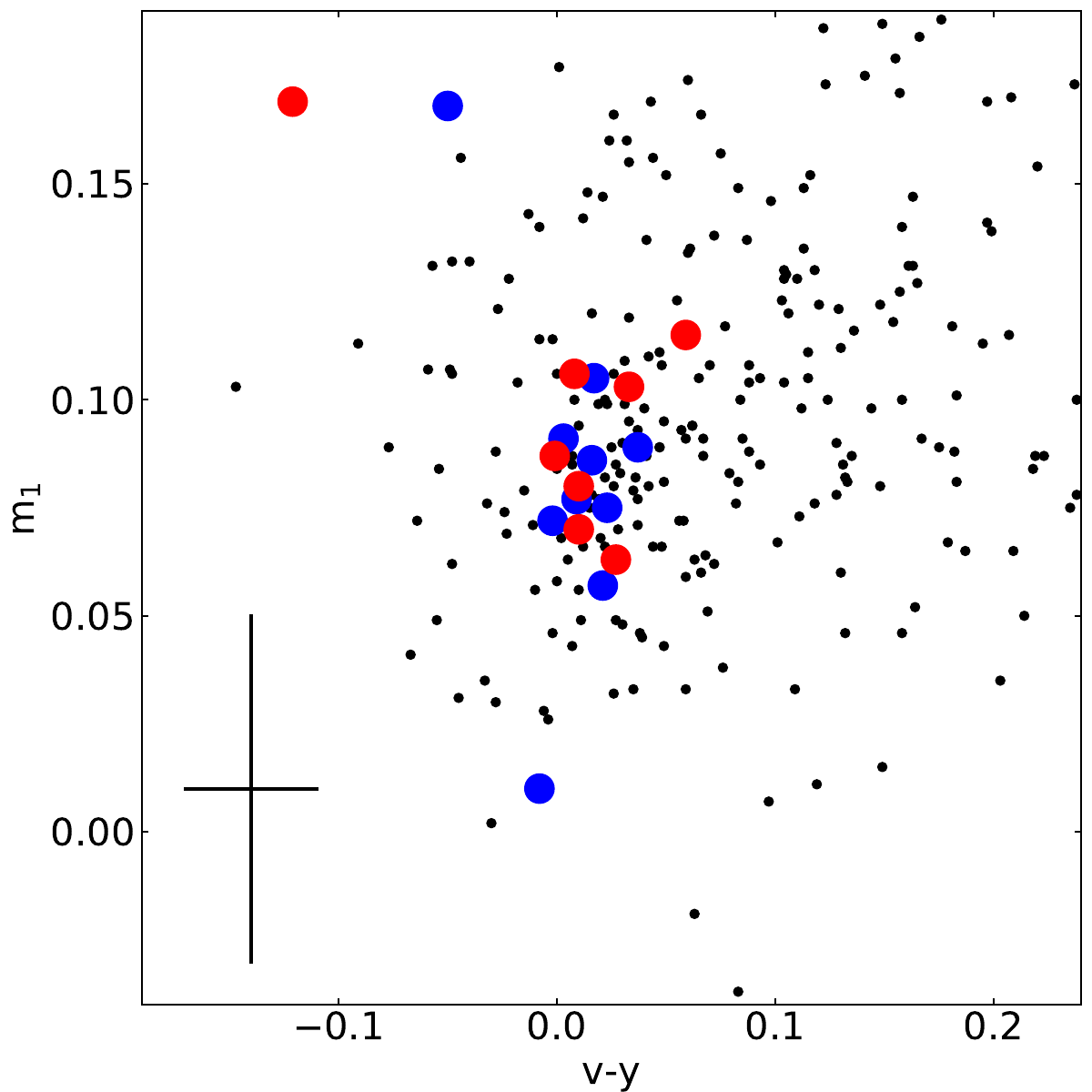}
\caption{Two-color diagram for stars measured in the field of NGC\,1971. Symbols as in Figure~\ref{fig:fig1}. Typical error bars for stars represented by blue and red filled circles
are also shown.}
\label{fig:fig2}
\end{figure}

The Str\"omgrem $m_1$ index is known as an iron abundance sensitive index 
\citep{calamidaetal2007}, which also measures variations in light-elements. Therefore, it might reflect
both heavy and light-element abundances or only those from light-elements. Because
metallicity variations were not found by \citet{pc2017}, we concluded that any distinction in the
$m_1$ values between blue and red stars could be caused by different CN/CH abundances, which 
in turn would imply the existente of two different populations among the stars in NGC\,1971.
Figure~\ref{fig:fig2} depicts the $m_1$ vs. $v-y$ diagram for all the measured stars in
Figure~\ref{fig:fig1}. It reveals that the two groups of stars selected from Figure~\ref{fig:fig1}
do not exhibit any visible separation. There is a couple of stars that remarkably
separate from the region where most of the blue and red ones are concentrated. These
stars could have anomalous chemical compositions, somehow defective Str\"omgren
photometric data, or be interlopers. Hence, Figure~\ref{fig:fig2} suggests that NGC\,1971 harbors
stars with similar light-element abundances. Blue and red stars also overlap in any color-color 
diagram that involves $b-y$, $v-y$ or $m_1$, which means that the Str\"omgren indices reveal
a single population young cluster with an homogeneous chemical abundance distribution.
The above results are expected for young star clusters, because MSTO stars have
higher temperatures as compared to those of old globular cluster stars.  In old globular
clusters the magnitude differences between CN-rich and CN-poor stars with similar luminosities are mostly due to the different strengths of molecules including light elements such us C, N or 
O \citep[e.g.,][]{marinoetal2008,yongetal2008,sbordoneetal2010}.  In hotter stars such differences are small, and are unlikely consistent with the large color difference between red and blue stars.

\begin{figure}
\includegraphics[width=\columnwidth]{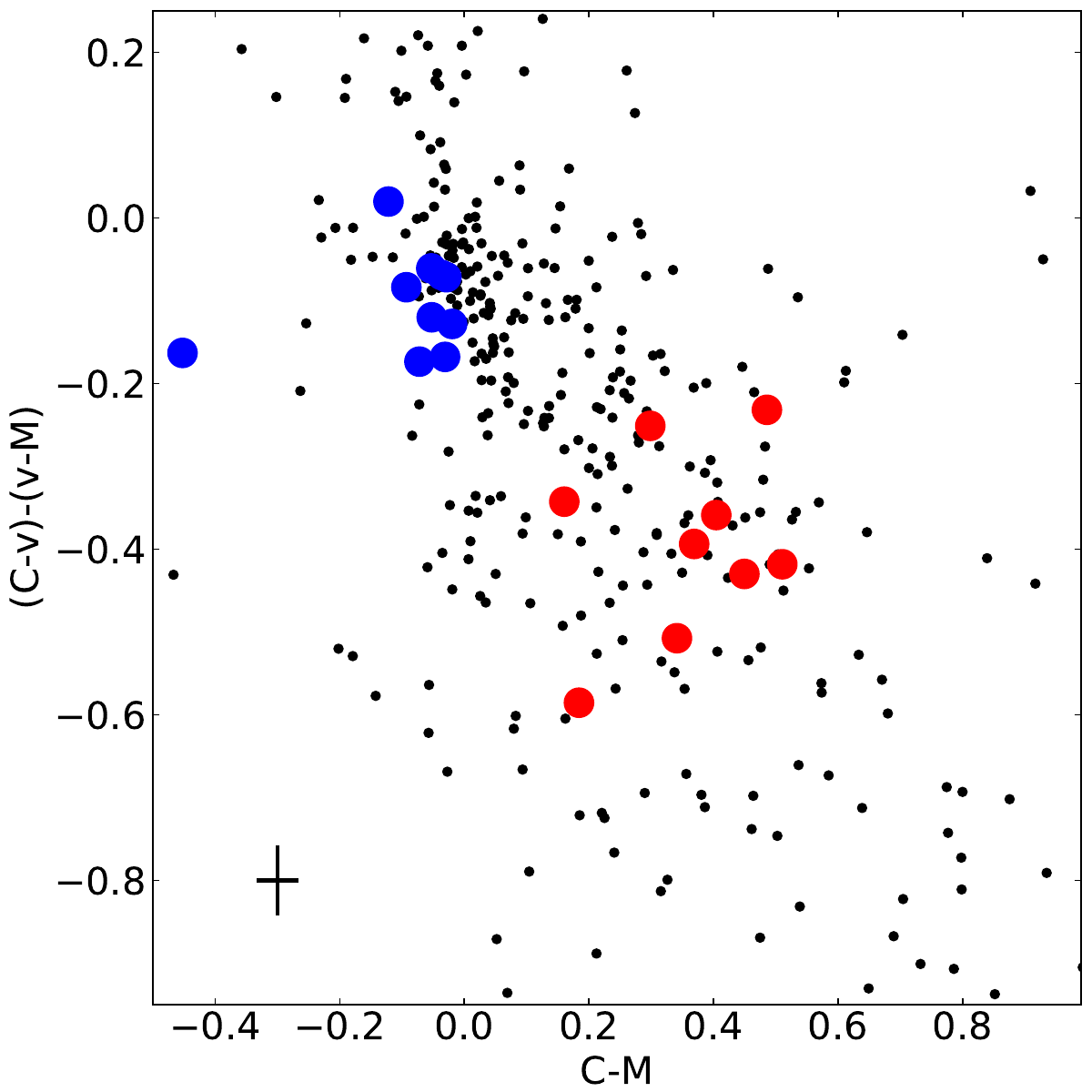}
\caption{Two-color diagram for stars measured in the field of NGC\,1971. Symbols as in Figure~\ref{fig:fig1}. Typical error bars for stars represented by blue and red filled circles
are also shown.}
\label{fig:fig3}
\end{figure}

\begin{figure}
\includegraphics[width=\columnwidth]{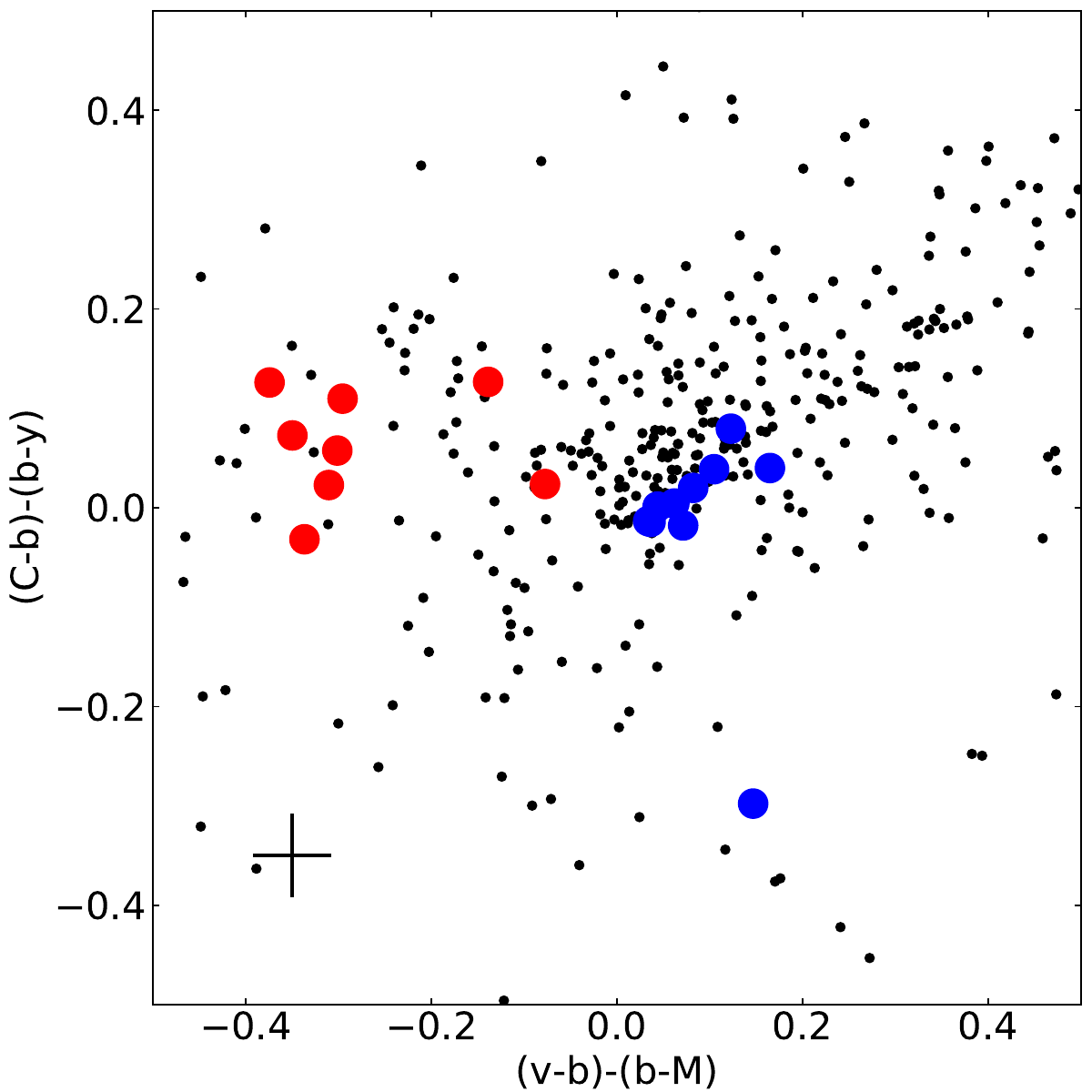}
\caption{Two-color diagram for stars measured in the field of NGC\,1971. Symbols as in Figure~\ref{fig:fig1}. Typical error bars for stars represented by blue and red filled circles
are also shown.}
\label{fig:fig4}
\end{figure}

Figure~\ref{fig:fig3} depicts the ($C-v$)-($v-M$) vs. $C-M$ diagram where the two groups of
stars selected from Figure~\ref{fig:fig1} are clearly differentiated. On the assumption that the
Washington-Str\"omgren metallicity sensitive passbands ($C,v$) are not measuring any
difference between blue and red stars, we concluded that the distinction between both
groups is caused by the Washington $M$ magnitudes. Indeed, it enters with a different
sign in the ordinate and abscissa of Figure~\ref{fig:fig3}. Moreover, we would arrive to the same
conclusion by comparing, for instance, $v-M$ vs $M-y$ with  $v-C$ vs $C-y$, among other
possible combinations. We confirmed such a behavior by plotting the pseudo color diagram 
($C-b$)-($b-y$) vs. ($v-b$)-($b-M$). In Figure~\ref{fig:fig4}, ($C-b$)-($b-y$) does 
not reflect any split between blue and red stars, as expected, while ($v-b$)-($b-M$) uncovers 
the different values due to the $M$ magnitude.

We searched the literature looking for spectral features of early MK-type stars 
in the wavelength range spanned by the Washington $M$ passband, and found that
the H$\beta$ emission line is a common feature of Be rotators
\citep[see, e.g.][]{miroshnichenkoetal2000}. According to
\citet{fangetal2018}, H$\beta$, H$\alpha$, and Ca\,II K emission lines are seen in
stars spanning a range of mass and rotation in open clusters as a consequence of their
chromospheric activity. They also found  that H$\beta$ and H$\alpha$ emissions are
correlated. 

At this point, we interpret that fainter and brighter $M$ magnitudes, responsible of the split 
between blue and red stars in Figures~\ref{fig:fig1}, \ref{fig:fig3}, and \ref{fig:fig4}, 
correspond to stars with fainter and stronger H$\beta$ emission lines, which in turn, refers to slow 
and fast rotators \citep{dmitrievetal2019}. Furthermore, the difference derived between
$M$ and $T_1$ magnitudes of blue and red stars (-0.26$\pm$0.07 mag) confirms
the correlation found by \citet{fangetal2018} between the H$\beta$ and H$\alpha$ emissions 
in open clusters.
The contribution of H$\beta$ to the $M$ magnitudes explains the eMSTO
observed by \citet{pc2017} in the $M$ vs. $C-M$ CMD (their figure 4), similarly as H$\alpha$
contributes to the $T_1$ magnitude in the $T_1$ vs. $C-T_1$ CMD. This outcome
shows that the eMSTO in the CMD of NGC\,1971 is observed whenever passbands
with Be emission line contributions are used. As far as we are aware, this is the first
time H$\beta$ emissions satisfactorily explain the existence of eMSTOs in young clusters.
Nevertheless, further spectroscopic confirmation is highly desired without a doubt. The
present outcome opens new possibilities of studying eMSTOs from observational and
theoretical perspectives, by using a larger number of photometric systems
with passbands centered on Be features, and by improving theoretical isochrones with
rotation effects that widen the range of rotation velocities considered.

\section{Conclusions}

The origin about the existence of eMSTOs in young star clusters is still under debate. During the last years,
it has been reached the general idea that neither a real age spread nor stellar evolutionary effects alone 
explain the observed broadness of the upper part of cluster MSs  \citep{goudfrooijetal2017,goudfrooijetal2018,gossageetal2019,lietal2020}.  Moreover, some star clusters do not seem to exhibit any eMSTO \citep{dejuanovelaretal2020}. In this context, 
the noticeable large age spread found for NGC\,1971 by \citet{pc2017}, which resulted to be similar to the
cluster age ($\sim$160 Myr), caught our attention.

We made use of publicly available and yet not exploited Str\"omgren $vby$ images centered on the 
cluster, obtained with the SOAR SOI, to search for additional clues about the noticeable eMSTO of
NGC\,1971. The accurate Str\"omgren photometry obtained, alongside with the published Washington
photometry, allowed us to explore the nature of the broadness at the upper part of the cluster MS
observed by \citet{pc2017}. During the analysis, we constrained the sample of stars to those with
 assigned  membership probabilities $P$ $\ge$ 75\% in both photometric data sets,
separately. The selected sample consists in 18 stars distributed from the top of the cluster MS down to
one magnitude underneath, and spans the range of eMSTO colors.

From an appropriate combination of Washington and Str\"omgren magnitudes we built different
pseudo colors and found that: i) the selected stars can be split into two groups of stars (called 'blue' and 'red'),
with 9 stars each; ii) the cluster does not seem to show inhomogeneities of light and heavy-element
abundances. On the contrary, both groups of selected stars resembles a single stellar population with 
an homogeneous chemical abundance level; iii) Blue and red stars are clearly separated only when 
the Washington $M$ magnitudes are involved, so that the $M$ magnitudes are responsible of the
appearance of the eMSTO. The spectral region covered by the $M$ passband comprises H$\beta$ which,
in emission, is a common feature of Be stars. Therefore, we speculate with the possibility that
Be stars populate the redder group of stars, in very good agreement with previous findings of
eMSTOs in young clusters \citep[e.g.][]{bastianetal2017}. Particularly, the present outcome reinforces
the suggestion that H$\alpha$ emissions of Be stars could have contributed to the $T_1$ magnitudes in
\citep{pc2017}, and hence to the detection of an eMSTO. In this work, we confirmed such a
possibility from the tight correlation between $M$ and $T_1$ magnitudes, as is also shown
between H$\beta$ and H$\alpha$ emissions of Be stars in open clusters.

 I thank the referee for the thorough reading of the manuscript and
the suggestions to improve it. 



\end{document}